\documentclass[10pt]{article}
\usepackage[dvipsnames]{xcolor}
\usepackage{geometry}
\usepackage{graphicx,caption}
\usepackage{floatrow}
\usepackage{bbm}
\usepackage[label font=bf,labelformat=simple]{subfig}
\newenvironment{Figure}
{\par\medskip\noindent\minipage{\linewidth}}  {\endminipage\par\medskip}
\usepackage{tikz}
\usepackage{amssymb,amsmath,amsthm}
\usepackage{hyperref}
\usepackage{polski}
\usepackage[utf8]{inputenc}
\usepackage[english]{babel}
\usepackage{multicol}
\usepackage{wrapfig}
\usepackage{xcolor} 
\usepackage{setspace} 
\usepackage{comment}
\usepackage[nolist]{acronym} 
\usepackage{orcidlink}
\usepackage{titlesec}
\titleformat*{\section}{\large\bfseries}
\titleformat*{\subsection}{\normalsize\bfseries}

\definecolor{cornellred}{rgb}{0.7, 0.11, 0.11}
\newcommand{\red}{\textcolor{cornellred}}

\hypersetup{
    pdfpagemode={UseOutlines},
    bookmarksopen,
    pdfstartview={FitH},
    colorlinks,
    linkcolor={RoyalBlue},
    citecolor={cornellred},
    urlcolor={ForestGreen}
            }
\usepackage{amsmath}
\DeclareMathOperator{\Tr}{Tr}

\title{\Large{ \bf Genus Comparisons in the Topological Analysis of
RNA Structures}}
 \author{Nicolò Cangiotti\footnote{Department of Mathematics, Politecnico di Milano,
via Bonardi 9, Campus Leonardo, 20133, Milan, Italy. Author to whom any correspondence should be addressed: \href{mailto:nicolo.cangiotti@polimi.it}{\texttt{nicolo.cangiotti@polimi.it}}.} \orcidlink{0000-0003-2200-446X}  \& \thinspace Stefano Grasso\footnote{Lesaffre International, 101 Rue de Menin, 59700 Marcq-en-Barœul, France.} \orcidlink{0000-0002-6958-9912}}
 \date{}

\begin{document}

\maketitle
\begin{acronym}[MPC] 
\acro{cWW}{Watson-Crick base-pairings}
\acro{ML}{machine learning}
\end{acronym}
\begin{abstract}
RNA folding prediction remains challenging, but can be also studied using a topological mathematical approach. In the present paper, the mathematical method to compute the topological classification of RNA structures and based on matrix field theory is shortly reviewed, as well as a computational software, McGenus, used for topological and folding predictions. Additionally, two types of analysis are performed: the prediction results from McGenus are compared with topological information extracted from experimentally-determined RNA structures, and the topology of RNA structures is investigated for biological significance, in both evolutionary and functional terms. Lastly, we advocate for more research efforts to be performed at intersection of physics-mathematics and biology, and in particular about the possible contributions that topology can provide to the study of RNA folding and structure. 
\end{abstract}

\vspace{15pt}

\begin{multicols}{2}
\section{Introduction}
\label{Sec1}
In the last decade, RNA science boomed, thus the range of its applications drastically increased, even more with the recent advent of synthetic biology \cite{Dykstra2022}. RNA can perform a diverse set of functions, which are already exploited for commercial applications, for instance it  modulates innate immunity, catalyzes chemical reactions, senses small molecules, and regulates gene expression \cite{Dykstra2022,Townshend2021}. RNA is a single-filament polymer, made of ribonucleotides with four main nucleobases (A, C, G, U), but that in reality can incorporate other bases. Despite being single-filament, and similarly to DNA, RNA bases can form bonds, with bases from other molecules, or more interestingly in this context, also within the same molecule, thus creating complex 2D and 3D structures. In the canonical \ac{cWW}, A binds U with two hydrogen bonds, and G binds C with three hydrogen bonds. As can be expected, in large proportion, what determines a RNA molecule's behaviour, is ultimately, its complex structure. RNA secondary structure is made of stable and fairly ``rigid'' conformations, which, in turns, fold hierarchically to form the tertiary structure, composed of more complex elements such as the \emph{pseudoknot}. A \emph{pseudoknot} is composed of at least two helices with internal bonds (\textit{e.g.}\ loops or hairpins) having at least a bond with each other \cite{Dam1992}. 

While RNA secondary structure is generally understood and can be reliably predicted \cite{Fallmann2017}, the same can not be said, despite the efforts, for RNA tertiary structure \cite{Townshend2021,Weeks2021}. However, to fully harness the potential of RNA molecules as biotechnological tools by engineering them, and because there is an intimate link between its structure and function, there is a increasing need to predict its folding. In this task, assistance can be provided by experimental data, which is still lacking at the moment and, unfortunately, remains hard to collect \cite{Dykstra2022,Townshend2021,Zhang2022}. This advocates for  more experimental efforts \cite{Dykstra2022, Townshend2021} and more understanding of the underlying physics or RNA folding. In fact novel \ac{ML} models are now able to combine large (experimental) data sets with physical constrains, in order to achieve better predictions \cite{Karniadakis2021}.

Interestingly, RNA folding can be approached by using some mathematical techniques coming from quantum mechanics. Indeed, we shall review how the identification of the contacts between nucleotides can be approached using the partition function of RNA folding in the combinatorial terms of \emph{Feynman diagrams}, and then translate it in terms of matrix field theory \cite{Orland2002}. It turns out that \emph{pseudoknots}, \textit{i.e.}\ the units composing the RNA tertiary structure, play a key role in RNA topological behaviour. 
In this context, the concept of \emph{RNA genus} was introduced exploiting the parallelism with mathematical topology and knot theory. 


In the present paper we aim at briefly summarizing some key concepts from the topological theory and evaluate its theoretical and practical applications towards the RNA molecule. We additionally question the biological meaning of the genus examining a dataset of RNA sequences, and, lastly, we advocate for more research to be carried out at the intersection of physical-mathematics and biology.

The paper is organized as follows. In Sect.\ \ref{Sec2} we retrace some of the mathematical and physical steps of the topological approach, introducing the diagrammatic notation and the key idea of genus. Sect.\ \ref{Sec3} is devoted to the description of the methods and the tools used for our analyses. Sect.\ \ref{Sec4} illustrates the results of: 1) the comparison of McGenus \cite{Bon12} \textit{ab-initio} topological predictions with experimentally determined ones;  2) within the analyzed dataset, the predicted relationships between taxonomy or RNA type and the topology of RNA are explored. Finally, in Sect.\ \ref{Sec5} we take stock of the work done, proposing intriguing research lines for the near future.

\section{Physical background}
\label{Sec2}
In 2002, Orland and Zee \cite{Orland2002} introduced a very intriguing method for predicting the tertiary structure of RNA sequences based on the so-called \emph{matrix field theory}. In particular, they noticed an analogy between the model obtained by stretching the classical manipulation of RNA secondary structure and the famous Feynman diagrams defined in quantum field theory and then also applied successfully in quantum chromdynamics (QCD). This approach is strictly connected with the topology of such diagrams, as proved by t'Hooft \cite{Hooft74}. The topological context allow us to introduce the notion of \emph{genus}, which provide the so-called topological classification of RNA structures \cite{Vernizzi04,Vernizzi16,Vernizzi06}. In this section, we briefly retrace some of the most significant mathematical and physical concepts of this predictive method, in order to better understand the transferability process from the domain of physics to the domain of biology. The energy models for RNA studied in recent years, are represented thanks to a so-called partition function, which in the three dimensional case, for a sequence of length $L$, is given by\footnote{Eqs. (1-8) are taken from \cite{Vernizzi04}.}
\begin{equation}
\label{eq:Zgeneral}
\mathcal{Z}=\int \prod_{k=1}^L \textrm{d}^3\mathbf{r}_k   f(\mathbf{r})Z_L(\mathbf{r}),
\end{equation}
where $\mathbf{r}_k$ is the 3D position vector $k$-th base, and $f(\mathbf{r})$ is a function, which takes into
account some properties of the RNA chain. The function $Z_L(\mathbf{r})$ provide the description of bases interactions and it is constructed as follows:
\begin{equation}
\scalebox{0.9}{$\displaystyle Z_L=1+\sum_{\langle i,j\rangle}V_{ij}(\mathbf{r}_{ij})+\sum_{\langle i,j,k,l\rangle}V_{ij}(\mathbf{r}_{ij})V_{kl}(\mathbf{r}_{kl})+\cdots$},
\end{equation}
where $\langle i,j\rangle$ denotes the pair relation $j>i$, $\langle i,j,k,l\rangle$ the quadruplets relation $l>k>j>i$, and so on. The function $V_{ij}(\mathbf{r})$ is the Boltzmann
factor with energy $\epsilon_{ij}$ that relates the $i$-th and the $j$-th base at the distance induced by $\mathbf{r}_{ij}$:
\begin{equation}
    V_{ij}(\mathbf{r}_{ij})=\exp \left(-\beta \epsilon_{ij}s_{ij}(\mathbf{r}_{ij})\right),
\end{equation}
where $\beta = 1/k_BT$ is the usual symbol for the inverse temperature multiplied to the Boltzmann constant, and $s_{ij}(\mathbf{r}_{ij})$ represents the space dependent part of the interaction. Assuming infinite flexibility for the chain, sterical constraints can be neglected. Therefore, one can drop all spatial degrees of freedom and get the simplified expression \cite{Orland2002,Vernizzi04}:
\begin{equation}
\label{eq:ZL} Z_L=1+\sum_{\langle i,j\rangle}V_{ij}+\sum_{\langle i,j,k,l\rangle}V_{ij}V_{kl}+\cdots,
\end{equation}
with
\begin{equation}
    V_{ij}=\exp \left(-\beta \epsilon_{ij}\right). 
\end{equation}
In the following lines, we briefly reviewed how the diagrammatic picture arises from mathematical theory. Moreover we highlight some important topological properties, which turn out to be of great interest in this biological framework, as extensively explained with different approaches in \cite{Andersen13a,Andersen13b,Bon08,Orland2002,Rubach2019,Vernizzi06,Xu20,Zajac2018}.

\subsection{The origin of the diagrams}
\begin{figure}[H]
\centering
\begin{tikzpicture}[scale=1,>=latex]
    \begin{scope}
    \draw[very thick,blue](1.41/2+1.5,1.41/2+1.5) to[out=240,in=120](1.41/2+1.5,-1.41/2+1.5);
     \draw[very thick,blue](-1.41/2+1.5,-1.41/2+1.5) to[out=60,in=300](-1.41/2+1.5,1.41/2+1.5);
    \draw (0,0)rectangle(6,6);
    \draw (0,3)--(6,3);
    \draw (3,0)--(3,6);
    \draw[black, thick] (1.5,1.5) circle[radius=1cm];
    \fill (1.41/2+1.5,1.41/2+1.5) circle[radius=2pt];
     \fill (-1.41/2+1.5,1.41/2+1.5) circle[radius=2pt];
     \fill (-1.41/2+1.5,-1.41/2+1.5) circle[radius=2pt];
     \fill (1.41/2+1.5,-1.41/2+1.5) circle[radius=2pt];
     \node[scale=0.75] at (1.5,0.25) {\textbf{c}};
    \end{scope}
    \begin{scope}[xshift=3cm]
     \draw[blue, very thick] (1.41/2+1.5,1.41/2+1.5)--(-1.41/2+1.5,-1.41/2+1.5);
     \draw[blue, very thick] (-1.41/2+1.5,1.41/2+1.5)--(1.41/2+1.5,-1.41/2+1.5);
     \draw[black, thick] (1.5,1.5) circle[radius=1cm];
     \fill (1.41/2+1.5,1.41/2+1.5) circle[radius=2pt];
     \fill (-1.41/2+1.5,1.41/2+1.5) circle[radius=2pt];
     \fill (-1.41/2+1.5,-1.41/2+1.5) circle[radius=2pt];
     \fill (1.41/2+1.5,-1.41/2+1.5) circle[radius=2pt];
     \node[scale=0.75] at (1.5,0.25) {\textbf{d}};
    \end{scope}
    \begin{scope}[xshift=3cm,yshift=3cm]
    \draw[very thick,blue](1.41/2+1.5,1.41/2+1.5) to[out=-150,in=-30](-1.41/2+1.5,1.41/2+1.5);
    \draw[black, thick] (1.5,1.5) circle[radius=1cm];
    \fill(1.41/2+1.5,1.41/2+1.5) circle[radius=2pt];
     \fill (-1.41/2+1.5,1.41/2+1.5) circle[radius=2pt];
     \fill (-1.41/2+1.5,-1.41/2+1.5) circle[radius=2pt];
     \fill (1.41/2+1.5,-1.41/2+1.5) circle[radius=2pt];
     \node[scale=0.75] at (1.5,0.25) {\textbf{b}};
    \end{scope}
    \begin{scope}[yshift=3cm]
     \draw[black, thick] (1.5,1.5) circle[radius=1cm];
     \fill (1.41/2+1.5,1.41/2+1.5) circle[radius=2pt];
     \fill (-1.41/2+1.5,1.41/2+1.5) circle[radius=2pt];
     \fill (-1.41/2+1.5,-1.41/2+1.5) circle[radius=2pt];
     \fill (1.41/2+1.5,-1.41/2+1.5) circle[radius=2pt];
     \node[scale=0.75] at (1.5,0.25) {\textbf{a}};
    \end{scope}
\end{tikzpicture}
\caption{Four examples of diagrams representing terms of the integral $Z_4(a)=1+6a+3a^2$. We represent here the no-chords diagram (\textbf{a}) that is the addendum $1$, a one-chord diagram (\textbf{b}) contributing to the first order term $6a$, and two of the three second order two-diagrams in $3a^2$: one planar (\textbf{c}) and one non-planar (\textbf{d}).}
\label{table}
\end{figure}
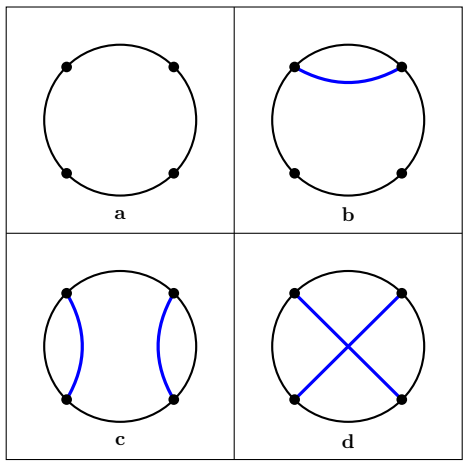

The fundamental role in the work of Orland and Zee \cite{Orland2002} is played by the following kind of integrals over the space of $N \times N$ dimensional Hermitian matrices:
\begin{equation}
\label{Eq:int_matrix}
\begin{aligned}
Z_n(a,N)=&\frac{1}{A(N)}\int \mathrm{d}^{N\times N}\phi \exp\left(-\frac{N}{2a}\right)\Tr{\phi^2}
\\
&\times \frac{1}{N}\Tr\left(\mathbbm{1}+\phi\right)^n,
\end{aligned}
\end{equation}
where the normalization factor $A(N)$ is given by
\begin{equation}
    \int \mathrm{d}^{N\times N}\phi \exp \left(-\frac{N}{2a}\Tr\phi^2\right).
\end{equation}
Here $\phi$ is an Hermitian matrix, namely it is equal to its transpose conjugate, and $\Tr(\cdot)$ represents the trace operator. To better understand the behaviour of such integrals and the link with the diagrams picture \cite{Bouttier11}, it could be helpful to consider the following one dimensional Gaussian integrals with the related normalization factor \cite{Vernizzi04}:
\begin{equation}
\begin{aligned}
Z_n(a)&=\frac{1}{A}\int_{\mathbb R} \mathrm{d}x \exp \left(-\frac{x^2}{2a}\right)(1+x)^n,\\
A&=\int_{\mathbb{R}} \mathrm{d}x \exp \left(-\frac{x^2}{2a}\right),
    \end{aligned}
\end{equation}
with the number $a\in\mathbb{R}_{>0}$ called, by analogy with the quantum theory, \emph{propagator}. It is now possible to provide a useful interpretation of these integrals in terms of diagrams, which, for historical reasons, are called \emph{Feynman diagrams}. If one takes, for instance, $n=2$, one immediately gets $Z_2(a)=1+a$, that is if one considers a circle with two points, the addendum $1$ stands for the so-called no chords diagram, and the terms $a$ stands for the diagram with now chords joining the two points. In the same way, it is possible to compute also $Z_4(a)=1+6a+3a^2$, where, in addition to the no-chords diagram, there are also $6$ one-chord diagrams (involving all the possible combinations) that represent the term $6a$. Finally, $3a^2$ are the $3$ possible combinations of two-chords diagrams. In Fig. \ref{table} we illustrate some examples of these diagrams. 

The power of this method turns out when one considers Eq. \eqref{Eq:int_matrix} for large value of $N$, namely the so-called large$-N$ expansion. Indeed, the computation of $Z_2(a,N)=1+a$ does not depend on $N$, but if one computes $Z_4(a,N)=1+6a+2a^2+a^2/N^2$ one gets explicitly the dependence on $N$ (as one can see by evaluating the above matrix integrals thank to the renowed Wick theorem). It is important to notice that the term involving $1/N^2$ represents, in diagrammatic terms the non-planar diagram: the one with the intersecting chords, that is diagram $(d)$ in Fig. \ref{table}. In \cite{Orland2002}, the authors exploit this analogy between Feynman diagrams and circle diagrams (see also Fig. \ref{kiss3}) used as representation of RNA secondary structure to study the possible pairings of RNA. Indeed, the higher-order terms in the expansions correspond to RNA secondary structures with pseudoknots. Thus, an in-depth analysis can help to tackle the problem of RNA-folding prediction, especially in the cases of lowest free energy.

In Sect.\ \ref{SecTopol}, we briefly explore the topological characterization of these graphs, similarly to that proposed for the the strong interaction of QCD by t'Hooft \cite{Hooft74}. In particular, we shall review an important topological number, the \emph{genus}, strictly correlated to the concept of pseudoknot \cite{Orland2002}. A more recent research about genus ranges of chord diagrams can be founded in \cite{Burns2015}.

\subsection{Genus and pseudoknots}
\label{SecTopol}
The usage of diagrams to describe RNA secondary and tertiary structure, as well as pseudoknots, is nowadays a practice, as testified by a large number of representations found in the literature as one can see, for instance, in \cite{Andersen13a,Andersen13b,Orland2002,Pillsbury05,Rubach2019,Vernizzi04,Vernizzi06,Xu20,Zajac2018}.

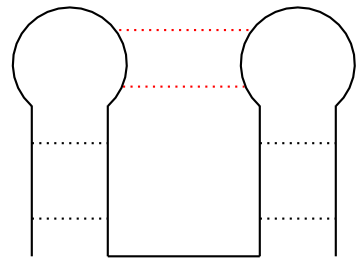
\begin{figure}[H]
\centering
\begin{tikzpicture}[scale=1,>=latex]
\draw[thick] (0,0)--(2,0);
\draw[thick] (0,0)--(0,2);
\draw[thick] (-1,0)--(-1,2);
\draw[thick] (3,0)--(3,2);
\draw[thick] (2,0)--(2,2);
\draw[thick,dotted] (-1,0.5)--(0,0.5);
\draw[thick,dotted] (-1,1.5)--(0,1.5);
\draw[thick,dotted] (2,0.5)--(3,0.5);
\draw[thick,dotted] (2,1.5)--(3,1.5);
\draw[thick,red,dotted] (0.2,2.25)--(1.8,2.25);
\draw[thick,red,dotted] (0.15,3)--(1.9,3);
\draw [thick,domain=-48:228] plot ({-0.5+0.75*cos(\x)}, {2.55+0.75*sin(\x)});
\draw [thick,domain=-48:228] plot ({2.5+0.75*cos(\x)}, {2.55+0.75*sin(\x)});
\end{tikzpicture}
\caption{Classical diagram of \emph{kissing hairpin} pseudoknot.}
\label{kiss1}
\end{figure}

\begin{figure}[H]
\centering
\begin{tikzpicture}[scale=0.8,>=latex]
\draw[thick] (-1,0)--(7,0);
\draw [very thick,dotted,domain=0:180] plot ({1.25+0.75*cos(\x)}, {0.75*sin(\x)});
\draw [very thick,dotted,domain=0:180] plot ({1.25+1.25*cos(\x)}, {1.25*sin(\x)});
\draw [very thick,dotted,domain=0:180] plot ({4.75+0.75*cos(\x)}, {0.75*sin(\x)});
\draw [very thick,dotted,domain=0:180] plot ({4.75+1.25*cos(\x)}, {1.25*sin(\x)});
\draw [very thick,dotted,red,domain=0:180] plot ({3+2*cos(\x)}, {2*sin(\x)});
\draw [very thick,dotted,red,domain=0:180] plot ({3+1.5*cos(\x)}, {1.5*sin(\x)});
\end{tikzpicture}
\caption{Stretching diagram of \emph{kissing hairpin} pseudoknot.}
\label{kiss2}
\end{figure}
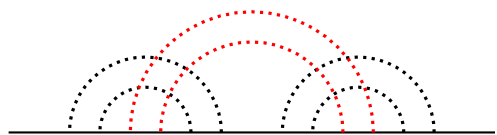

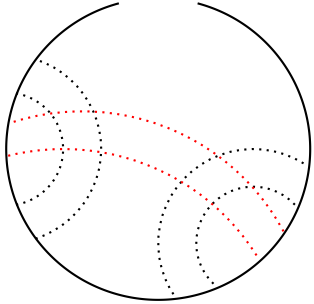
\begin{figure}[H]
\centering
\begin{tikzpicture}[scale=1,>=latex]
\draw [thick,samples = 2000,domain=105:435] plot ({2+2*cos(\x)}, {2+2*sin(\x)});
\draw [thick,dotted,domain=-80:81] plot ({0+0.75*cos(\x)}, {2+0.75*sin(\x)});
\draw [thick,dotted,domain=-72:72] plot ({0+1.25*cos(\x)}, {2+1.25*sin(\x)});
\draw [thick,dotted,domain=43:228] plot ({3.25+0.75*cos(\x)}, {0.75+0.75*sin(\x)});
\draw [thick,dotted,domain=58:215] plot ({3.25+1.25*cos(\x)}, {0.75+1.25*sin(\x)});
\draw [thick,dotted,red,domain=28:108] plot ({1+3*cos(\x)}, {-0.5+3*sin(\x)});
\draw [thick,dotted,red,domain=32:105] plot ({0.75+3*cos(\x)}, {-1+3*sin(\x)});
\end{tikzpicture}
\caption{Circle diagram of \emph{kissing hairpin} pseudoknot.}
\label{kiss3}
\end{figure}

Starting from the classical representation of an RNA molecule and its interaction, \textit{e.g.}\ the \emph{kissing hairpin} in Fig. \ref{kiss1}, it is possible, by stretching the backbone to obtain other two useful depictions, namely the \emph{stretching} (or \emph{arc}) \emph{diagram} (Fig. \ref{kiss2}) and the \emph{circle} (or \emph{disk}) \emph{diagram} (Fig. \ref{kiss3}).

It is important to underline that the lines associated to the base pairings have to belong to the \emph{same side} of the diagram (the upper or lower side for the stretching diagram and inside or outside of the disk for the circle diagram). The aim of this section is to highlight the topological connection between crossing diagrams and pseudoknots. A common way to face such a problem is to consider the diagram drawn by using the so-called \emph{double line notation} as illustrated in \cite{Orland2002}. However, for the purpose of this work, it is sufficient to grasp the intuitive underlying idea. 

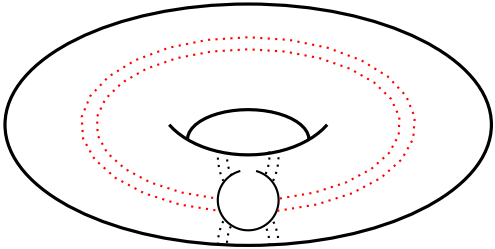
\begin{figure}[H]
\centering
\begin{tikzpicture}[scale=0.8,>=latex]
\draw[very thick] (2,2) ellipse (4cm and 2cm);
\draw [very thick,domain=0:180] plot ({2+1*cos(\x)}, {1.75+0.5*sin(\x)});
\draw [very thick,domain=210:330] plot ({2+1.5*cos(\x)}, {2.5+1*sin(\x)});
\draw [thick,samples = 1000,domain=105:435] plot ({2+0.5*cos(\x)}, {0.75+0.5*sin(\x)});
\draw [thick,dotted,red,domain=-80:260] plot ({2+2.75*cos(\x)}, {2+1.45*sin(\x)});
\draw [thick,dotted,red, domain=-78:258] plot ({2+2.5*cos(\x)}, {2+1.25*sin(\x)});
\draw [thick,dotted, domain=174:215] plot ({2+0.5*cos(\x)}, {1.5+0.75*sin(\x)});
\draw [thick,dotted, domain=-35:6] plot ({2+0.5*cos(\x)}, {1.5+0.75*sin(\x)});
\draw [thick,dotted, domain=175:217] plot ({2+0.35*cos(\x)}, {1.5+0.55*sin(\x)});
\draw [thick,dotted, domain=-35:6] plot ({2+0.35*cos(\x)}, {1.5+0.55*sin(\x)});
\draw [thick,dotted, domain=0:40] plot ({2+0.35*cos(\x)}, {0.55*sin(\x)});
\draw [thick,dotted, domain=0:40] plot ({2+0.5*cos(\x)}, {0.7*sin(\x)});
\draw [thick,dotted, domain=145:180] plot ({2+0.35*cos(\x)}, {0.55*sin(\x)});
\draw [thick,dotted, domain=145:180] plot ({2+0.5*cos(\x)}, {0.7*sin(\x)});
\end{tikzpicture}
\caption{\emph{Kissing hairpin} pseudoknot embedding on a
torus. We notice that the circle diagram can be actually drawn without any crossings. This corresponds to the topological genus of the torus, namely $g = 1$.}
\label{kiss_toro}
\end{figure}

The pseudoknots characterization in terms of crossing diagrams is well expressed by the \emph{genus} of a surface. Topologically speaking, the genus of a surface is the number number of \emph{holes} and \emph{handles} of a (orientable) surface. In this framework, the genus of diagram can be defined as the genus of the surface with the lowest genus (\textbf{i.e.}\ number of \emph{holes}) in which our diagram can be drawn without intersections. In the case of the kissing hairpin, as one can see in Fig. \ref{kiss_toro} (adapted from \cite{vernizzi04pp}), this is possible in a surface with genus $g=1$, namely a torus.

This topological picture allows us to classify pseudoknots by rewriting the series expansion defined in the previous section. Indeed, in \cite{vernizzi04pp,Vernizzi06} the genus g is included explicitly from the original model \cite{Orland2002}, by considering power series with respect terms of the form $N^{-2g}$ (here $N$ denotes, as above, the dimension of the matrix). A practical way to compute the genus of a diagram comes from the celebrated Euler characteristic that in the case of diagrams is defined as $\chi=V-E+F$, where $V,E$, and $F$ are the numbers of vertices, edges, and faces respectively. In this description, a vertex is a nucleotide, an edge is any line connecting two nucleotides, and a face is a part of the surface within a closed loop of edges. In the case of $n$ arcs one trivially $E = V + n$. Moreover, there is a famous theorem due to Euler  stating that any polyhedron homeomorphic to a sphere with a boundary has an Euler characteristic $\chi = 1$. As a corollary, all RNA secondary structures without pseudoknots are represented by disk diagrams with $\chi = 1$. Let us suppose that the RNA secondary structure admits pseudoknots, as in the case of kissing hairpin, the computation of the Euler characteristic leads\footnote{Sometimes, for computing correctly the values of $V,E$, and $F$, is useful to employ the practical \emph{double line notation} already mentioned.} to the value $\chi=-1$. The geometrical significance of such a value is strictly related to the number of \emph{holes} (or \emph{handles}) of a surface. In particular, for an orientable surface we have $\chi=2-2g-p$, where $p$ is the number of punctures (clearly $p=1$ for the single RNA stand). Thus, the kissing hairpin pseudoknot induces a genus $g=1$, and can be drawn without crossing on a surface with one \emph{hole}, that is, as said before, a \emph{torus} (see Fig. \ref{kiss_toro}). 

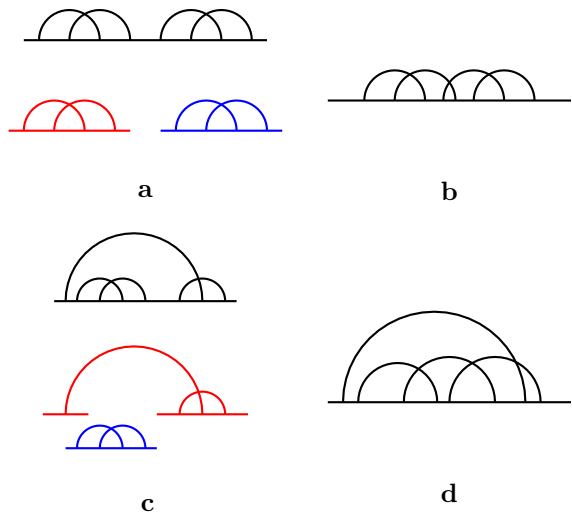
\begin{figure}[H]
\centering
\begin{tikzpicture}[scale=0.8,>=latex]
    \begin{scope}[xshift=0.625cm,yshift=0.3cm,scale=.75]
    \draw[thick] (0.5,4.5)--(4.5,4.5);
     \draw[thick,red] (0.25,2)--(1.25,2);
     \draw[thick,red] (2.75,2)--(4.75,2);  
     \draw[thick,blue] (0.75,1.25)--(2.75,1.25); 
    \draw [thick,domain=0:180,red] plot ({3.75+0.5*cos(\x)}, {2+0.5*sin(\x)});
     \draw [thick,domain=0:180,red] plot ({2.25+1.5*cos(\x)}, {2+1.5*sin(\x)});
      \draw [thick,domain=0:180,blue] plot ({1.5+0.5*cos(\x)}, {1.25+0.5*sin(\x)});
      \draw [thick,domain=0:180,blue] plot ({2+0.5*cos(\x)}, {1.25+0.5*sin(\x)});
     \draw [thick,domain=0:180] plot ({1.5+0.5*cos(\x)}, {4.5+0.5*sin(\x)});
     \draw [thick,domain=0:180] plot ({2+0.5*cos(\x)}, {4.5+0.5*sin(\x)});
     \draw [thick,domain=0:180] plot ({2.25+1.5*cos(\x)}, {4.5+1.5*sin(\x)});
     \draw [thick,domain=0:180] plot ({3.75+0.5*cos(\x)}, {4.5+0.5*sin(\x)});
      \node[scale=1] at (2.55,0) {\textbf{c}};
    \end{scope}
    \begin{scope}[xshift=5cm]
     \draw[thick] (0.5,2)--(4.5,2);
     \draw [thick,domain=0:180] plot ({2.25+1.5*cos(\x)}, {2+1.5*sin(\x)});
     \draw [thick,domain=0:180] plot ({3.25+0.75*cos(\x)}, {2+0.75*sin(\x)});
      \draw [thick,domain=0:180] plot ({2.5+0.75*cos(\x)}, {2+0.75*sin(\x)});
      \draw [thick,domain=0:180] plot ({1.65+0.65*cos(\x)}, {2+0.65*sin(\x)});
     \node[scale=1] at (2.5,0.5) {\textbf{d}};
    \end{scope}
    \begin{scope}[xshift=5cm,yshift=5cm]
    \draw[thick] (0.5,2)--(4.5,2);
    \draw [thick,domain=0:180] plot ({1.6+0.5*cos(\x)}, {2+0.5*sin(\x)});
     \draw [thick,domain=0:180] plot ({2.1+0.5*cos(\x)}, {2+0.5*sin(\x)});
     \draw [thick,domain=0:180] plot ({2.9+0.5*cos(\x)}, {2+0.5*sin(\x)});
     \draw [thick,domain=0:180] plot ({3.4+0.5*cos(\x)}, {2+0.5*sin(\x)});
      \node[scale=1] at (2.5,0.5) {\textbf{b}};
    \end{scope}
    \begin{scope}[yshift=5cm]
     \draw[thick] (0.5,3)--(4.5,3);
     \draw[thick,red] (0.25,1.5)--(2.25,1.5);
     \draw[thick,blue] (2.75,1.5)--(4.75,1.5);  
     \draw [thick,domain=0:180,red] plot ({1+0.5*cos(\x)}, {1.5+0.5*sin(\x)});
      \draw [thick,domain=0:180,red] plot ({1.5+0.5*cos(\x)}, {1.5+0.5*sin(\x)});
      \draw [thick,domain=0:180,blue] plot ({3.5+0.5*cos(\x)}, {1.5+0.5*sin(\x)});
      \draw [thick,domain=0:180,blue] plot ({4+0.5*cos(\x)}, {1.5+0.5*sin(\x)});
     \draw [thick,domain=0:180] plot ({1.25+0.5*cos(\x)}, {3+0.5*sin(\x)});
     \draw [thick,domain=0:180] plot ({1.75+0.5*cos(\x)}, {3+0.5*sin(\x)});
     \draw [thick,domain=0:180] plot ({3.25+0.5*cos(\x)}, {3+0.5*sin(\x)});
     \draw [thick,domain=0:180] plot ({3.75+0.5*cos(\x)}, {3+0.5*sin(\x)});
      \node[scale=1] at (2.5,0.5) {\textbf{a}};
    \end{scope}
\end{tikzpicture}
\caption{Four example of pseudoknot. A reducible (\textbf{a}) and an irreducible (\textbf{b}) pseudoknot. We can split (\textbf{a}) in two disconnected pieces with a single cut. A nested (\textbf{c}) and a non-nested (\textbf{d}) pseudoknot. We can obtain two disconnected components from (\textbf{c}) by making two cuts.}
\label{table2}
\end{figure}

Finally, we want to stress two more properties of pseudoknots which are very important for the realization of the software McGenus \cite{Bon12} used in this work. Indeed, the genus of a diagram is an additive quantity, an so it is possible to provide two notions to characterize the intrinsic complexity of a pseudoknot, namely the concepts of \emph{irreducibility} and \emph{nested} pseudoknot \cite{Bon08}. A diagram is said to be irreducible if it cannot be split into two disconnected parts by cutting a single line, as in Fig. \ref{table2} (\textbf{b}). In parallel, a diagram is said to be nested in another one if it can be removed by cutting two lines and keeping the rest of the diagram connected in a single component, as in Fig. \ref{table2} (\textbf{d}). These two definitions can be combined as follows: if a diagram is both irreducible and non-nested is called a \emph{primitive} diagram. The interested reader can find a more detailed investigations on genus and pseudoknot in this context in, \textit{e.g.}\   \cite{Bon10,Bon08,Vernizzi04,Vernizzi16,Vernizzi06}.

In the recent years, two interesting generalization of the concept of genus were introduced \cite{Rubach2019,Zajac2018}, which we summarize for the sake of completeness. The first one is the so-called \emph{genus trace}, a function $g(i):\mathbb{N}\rightarrow\mathbb{N}$ providing the genus of a segment of the chain between the first and the $i$-th residue. The second one is instead the \emph{fingerprint matrix}, which gives a useful mathematical visualization of all the genuses computed between two elements of a chain, namely if one uses the notation $\mathbf{G}=(g_{ij})$ for the matrix, the generic element $g_{ij}$ represent the genus of the sub-chain between the $i$-th and the $j$-th residue.

Thanks to the topological interpretation and some considerations around statistical mechanics model of RNA, a powerful algorithm has been proposed in \cite{Bon12,Bon11}, as we are going to see in the next part of the paper.

\section{Methods}
\label{Sec3}
In the previous section we have reviewed a topological description strictly related to the so-called large-$N$ matrix field theory. Such a framework has been adapted in \cite{Bon12,Bon11,vernizzi04pp} to develop a software for modeling RNA structure. Indeed, it is possible to rewrite the partition function from Eq.\ \eqref{eq:Zgeneral} with a more comfortable notation, highlighting some fundamental terms involved in the free energy contribution. In particular, following \cite{vernizzi04pp}, one considers the following partition function
\begin{equation}
\label{eq:Znew}
\mathcal{Z}=\sum_{C_{ij}}\exp\left[-\frac{1}{k_BT}\left(E(C)+T\cdot S(T,C)\right)\right],
\end{equation}
where $E(C)$ is the energetic contribution, $S(C,T)$ is the entropic contribution, and $C$ is the $L\times L$ \emph{contact matrix}, whose elements are
\[
C_{ij}=\begin{cases}
1 &\text{if $i$ and $j$ are paired};\\
0 &\text{if $i$ and $j$ are not paired}.
\end{cases}
\] 
Is important to notice that the terms involved in the Eq.\ \eqref{eq:Znew} have been determined empirically and they are called the ``Turner energy rules'' \cite{Serra95,Zuker99}. The contribution of the topological fluctuations due to the pseudoknots, as the kissing hairpin mentioned in Sect.\ \ref{SecTopol}, can be involved into the equation as follows
\[
\mathcal{Z}=\sum_{C_{ij}}\exp\left[-\frac{1}{k_BT}\left(E(C)+T\cdot S(T,C)+\mu\cdot g(C)\right)\right],
\]
where $\mu$ stands for the \emph{topological chemical potential} and $g(C)$ is the genus of the configuration associated to the contact matrix $C$. The relation between $\mu$ and the order $N$ of the matrix formulation provided in Sect.\ \ref{Sec2} is given by the following equation \cite{vernizzi04pp}
\[
\mu=-2k_BT\log(N).
\]

In the next paragraphs, we briefly summarize the main features and techniques employed in our analysis. The focus is on the software McGenus \cite{Bon12}, which is based on the topological characterization described above, and on the ‘Genus for biomolecules’ database \cite{Rubach2019}, from which we have extracted the experimental RNA data for the comparison, as described in Sect.\ \ref{dataset}, whose results are presented in Sect.\ \ref{Sec4}.

\subsection{Dataset} 
\label{dataset}
\emph{Genus} data (trace and total genus, the latter used in our comparisons) were downloaded from the ‘Genus for biomolecules’ database (\url{http://genus.fuw.edu.pl}) \cite{Rubach2019}.
In the ‘Genus for biomolecules’ database $1575$ unique RNA structures used in \cite{Zajac2018} to compute the genus and its derivatives were present. Of those $1575$ unique RNA structures, we discarded those whose sequence was not unique (an RNA sequence can be associated with multiple structures, \textit{e.g.}\ different conditions) and those with length higher than $1000$ nucleotides, remaining with $739$ unique sequences.  Additionally, we used sequences as reported in their canonical form (\textit{i.e.}\ only A, C, G, U), thus ignoring the many non-canonical bases incorporated in such molecules.

We further expanded this dataset with an additional $408$ RNA sequences defined as those deposited in RNAcentral release $22$ \cite{Sweeney2020}, and associated with a PDB identifier. Sequences were thus downloaded from PDB \cite{Berman2000a}. Because of the technical limitations of McGenus, outlined in Sect.\ \ref{mcgen}, we excluded sequences longer than $1000$ nucleotides. 
Through this part of the dataset, we performed a nucleotide replacement, substituting N, X, F, and M with A, and substituting V with G.

For the whole dataset, thus consisting of $1147$ unique sequences, taxonomical and RNA type metadata were retrieved using the EBI search API \cite{Madeira2022}; taxid numbers were then converted into respective names using the python package TaxidTools v.\ 2.2.3 \cite{Denay2021}. When TaxidTools was not able to retreive the taxonomic information associated with a sequences, in most cases due to synthetic constructs, the label ``unknown'' has been assigned to all taxonomical information. The whole dataset is available at: \url{https://github.com/grassoste/genus-comparisons}. 

\subsection{McGenus}
\label{mcgen}
We based our comparative analysis on the McGenus software (v7.0, August 2011), in turns rooted on algorithms that exploit the theoretical framework presented in the previous sections. Such a software is extensively described in \cite{Bon12} and it represents an important development of its precursor TT2NE \cite{Bon11}. Both McGenus and TT2NE start from the same energy function , which is defined in terms of helipoints minimazing the free energy. In both cases it is added the genus
penalty defined above, so that the free energy $F_S$ of a given RNA structure $S$ can be written as
\[
F_S=\sum_{i=1}^N\sigma_i^S\cdot\Delta F(h_i)+\mu\cdot g(S),
\]
where $\sigma_i^S$ is a sort of indicator function that takes on the value $0$ or $1$ whether helipoint $h_i$ belongs to $S$, $\Delta F$ is a local term indicating the free energy of individual helipoint and $g$ is the global terms representing the topological genus presented in the previous section, as well as the penalizing parameter $\mu$.
However, 
while TT2NE uses a deterministic order for adding or removing helipoints in the construction of the secondary structure, McGenus adds or removes one at a time based on a stochastic Monte Carlo (MC) Metropolis
scheme. We want to underline that there is no restriction on the pseudoknot topology that McGenus can generate. 

The main limitations of these 2 tools are the length of admitted sequences, up to $1000$ bases for McGenus and up to $200$ bases for TT2NE. And the fact that can only account for the canonical \ac{cWW}, thus underestimating the possible bonds. Further information on the algorithms can be found in \cite{Bon12,Bon11}. For such reasons the dataset had been generated and curated as described in Sect.\ \ref{dataset}.

We run the software with the option \texttt{-nsuboptimal} set to $10$ and the \texttt{-maxgenus} set to a $+5$ compared to the total genus trace as calculated by \cite{Rubach2019} for the initial $739$ sequences, and set to $5$ for the remaining $408$ sequences. Out of these, $30$ showed an average genus (\textit{i.e.}\ the average genus of the $10$ sub-optimal structures) of $4$ or higher and were thus re-run with a \texttt{-maxgenus} of $15$. 

\subsection{Code and analyses}
McGenus wrapper scripts were developed in-house and written in Python. Because McGenus treats RNA structures as statistical ensembles, and the best 10 structures are returned, we decided to use the mean of the 10 structures. Obviously non-integer numbers of genus do not have a counterpart in the mathematical realm, yet we believe that in this way we better capture the molecular dynamism of RNA sequences. Calculations of the parameters of the regression equations were performed with the Python package \texttt{scipy} v.\ 1.8.0.

\section{Results and discussion}
\label{Sec4}
In order to evaluate the usefulness of the idea of genus, borrowed from the realm of mathematical topology and applied to biological molecules, specifically RNA, we analyzed a set of $1147$ RNA sequences. Sequences were phylogenetically diverse and belonging to different types of RNA molecules.

\subsection{McGenus software}
\label{Mc_results}
Before presenting and discussing the actual results, a brief premise regarding the tool used is necessary. McGenus is a powerful tool and one of the few trying to predict the tertiary structure using, as input, only an RNA sequence in text format. Nevertheless, also due to its age, its running time for long sequences is high, despite being written in C to have a higher efficiency of computation.  Additionally, as already explained in Sect.\ \ref{Sec3}, it poses some technical limitations related to the length and nucleotide composition of molecules, thus preventing the analysis of some functionally interesting RNA molecules, for example long ribosomal RNA (rRNA) or long non-conding RNA (lncRNA) sequences. Lastly, its functionality is a bit limited, not allowing more than a sequence at a time nor for parallel computations, thus needing the development of python or R wrappers for a more efficient data handling. 

Despite the limitations, in addition to using McGenus to calculate the (total) genus of RNA molecules, we were able to show that McGenus can even be used to calculate the respective genus trace. We showed this concept for the RNA sequence with PDB ID 6P5N (chain 1), which is also present in the ‘Genus for biomolecules’ database (genus trace can be found at \url{https://genus.fuw.edu.pl/view/6p5n/1/}). Differently from the genus trace calculated from a 3D structure, the genus trace obtained from McGenus is composed of statistical ensembles, thus having an higher degree of noise as shown in Fig. \ref{genus_trace}. Yet, the similarity with the structure-based genus trace is striking, clearly showing the two main domains of this molecule. Additionally, McGenus provides extra information, based on the variability of the statistical ensemble, an indirect measure of its structural stability. It is thus clear that  McGenus could be exploited also in sight of the most recent development of the topological concept of genus applied to RNA molecules.

Novel implementations of McGenus algorithm, or alternative ones with the same functions, that circumvent the above mentioned limitations, would facilitate the exploration and exploitation of the genus concept in biology. This could have potentially far-reaching consequences in the development of novel \ac{ML} tools for RNA tertiary structure prediction, or in the design of engineered RNA molecules, or simply in the understating of the complex structure-function relation within nucleic acids. 

\begin{Figure}
 \centering
\includegraphics[width=0.95\linewidth]{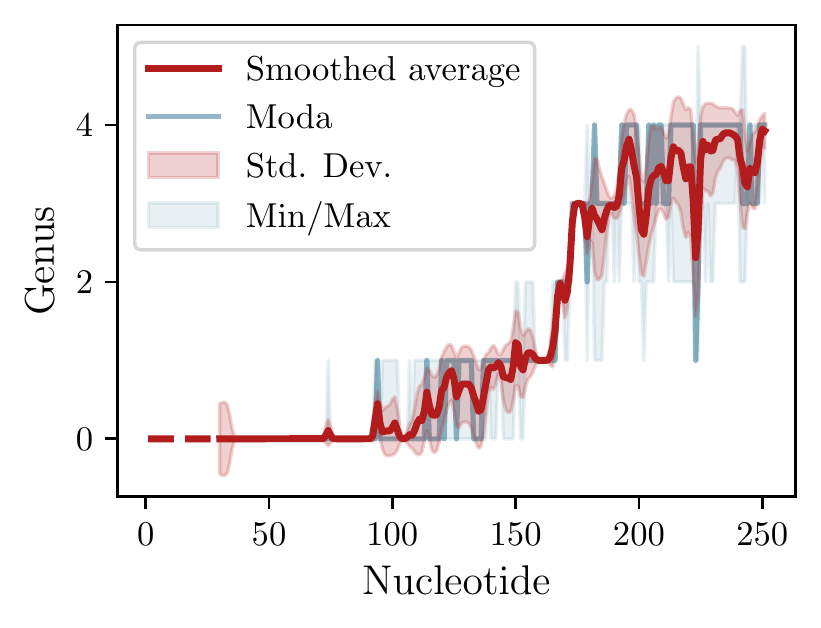}
\captionof{figure}{Genus trace of RNA with PDB ID 6P5N chain 1 predicted using McGenus. Average and mode of the 10 suboptimal structures of the statistical ensemble are reported. The same values have been used to shade the area included in the standard deviation (Std.\ Dev.)\ area and between the respective minimum and maximum (Min/Max) values. The two main domains of this molecule are clearly visible between nucleotides 100-150 and 200-250. Areas of higher variability are an indirect representation of higher structural variability in such area. Average and Std.\ Dev.\ have been smoothed for better representation. Dashed line indicates missing data.}
 \label{genus_trace}
\end{Figure}

\subsection{Experimental or computational genus?}
\label{Sec41}

Our dataset consisted in mostly relatively short RNA sequences (see Fig. \ref{gen_dist}). In parallel, the genus predicted for these sequences was on average quite low, with only longer sequences having a higher genus (see Fig. \ref{gen_dist} and Fig. \ref{fig:linreg} (\textbf{a})). Similarly to \cite{Zajac2018}, we also found a consistent relationship between a sequence length and its genus, albeit, even considering only the canonical \ac{cWW}, the difference is more than two folds. Regressing the average genus of the 10 suboptimal structures with the respective sequence length yielded that $g = 0.0134d$, where $d$ denotes the length of the sequence in nucleotides (see the black line in Fig. \ref{fig:linreg} (\textbf{a}) and its equation). Interestingly using the total genus, as calculated in \cite{Zajac2018} for the sequences available for the analysis with McGenus, for the regression, results in $g = 0.0529d$, slightly less than reported in the original publication, while the slope remains unchanged using McGenus results. Because we do not have the total genus calculated using only the canonical \ac{cWW} for the common subset, we could assume the regression value would be slightly lower, probably due to the lack of the longer sequences, increasing the difference. This would mean that somehow McGenus predictions overestimate the genus when only considering the canonical \ac{cWW}. Nevertheless, looking at the global picture, it is clear that non canonical pairings play an important role in RNA tertiary structure, and not taking them into account is a major limitation.

In both cases, some sequences present a genus particularly higher or lower, considering their length. It would be interesting to understand if there is a biological significance from these deviations from the norm, as it has been demonstrated to be the case for other biological relations \cite{Fudenberg2012}.

\begin{Figure}
 \centering
\includegraphics[width=\linewidth]{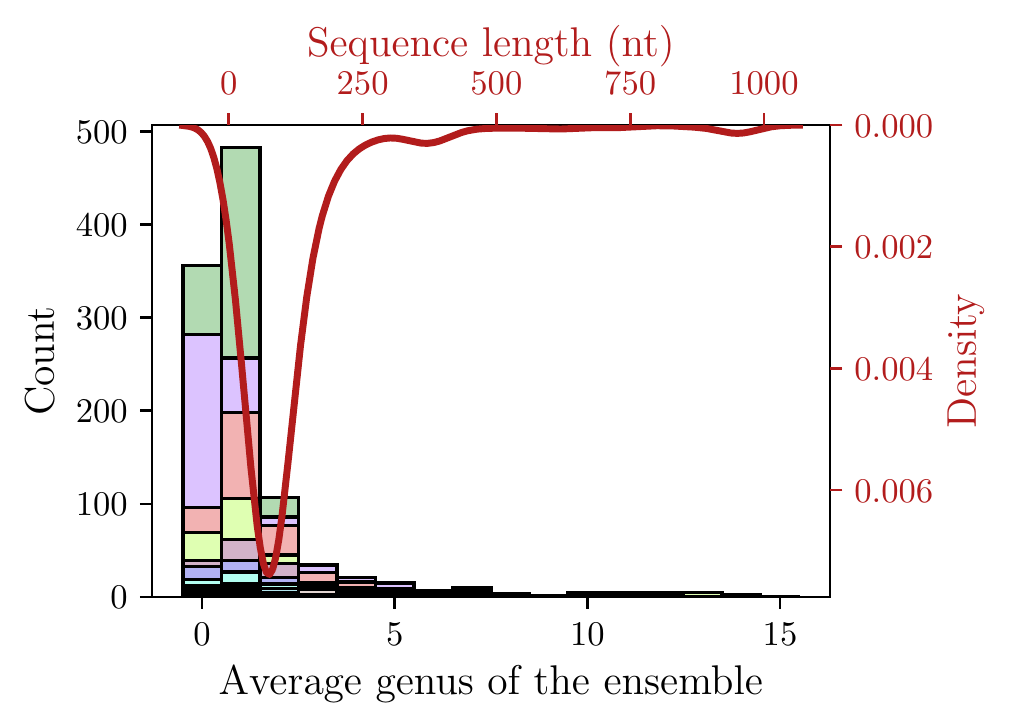}
\captionof{figure}{Distribution of sequence lengths and the genus values obtained in our analysis. Results show how in the analyzed dataset most structures are predicted to have a low genus value. This could be due to the majority of sequences being relatively short. Legend has been omitted for sake of space, different RNA types are colored as is Fig. \ref{fig:linreg} (\textbf{a}). }
 \label{gen_dist}
\end{Figure}



\subsection{Exploring genus (dis)homogeneity}
\label{Sec42}
\begin{figure*}[p!]
     \centering
     \sidesubfloat[]{\includegraphics[width=0.95\textwidth]{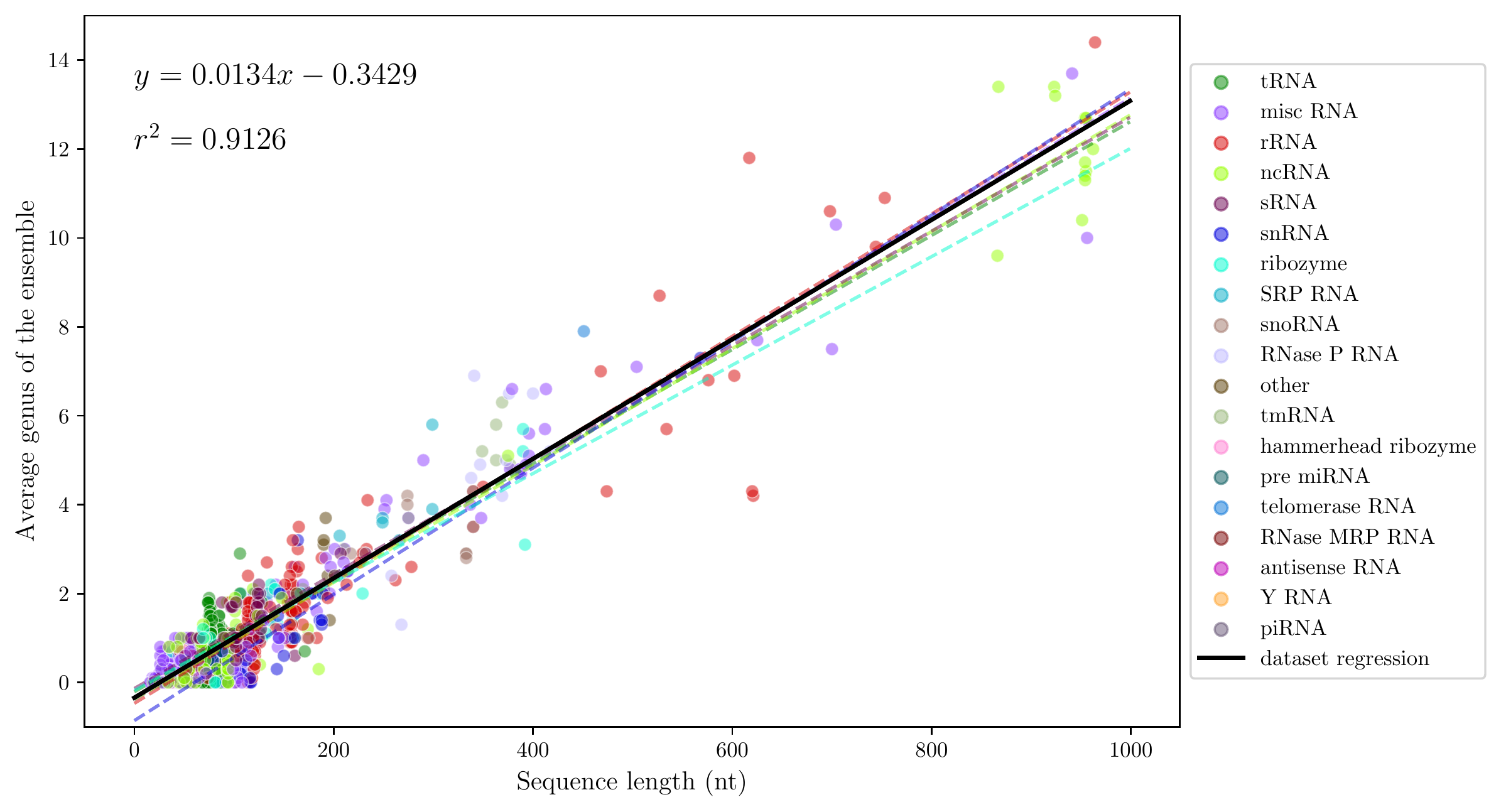}}\\[15pt]
     \sidesubfloat[]{\includegraphics[width=0.95\textwidth]{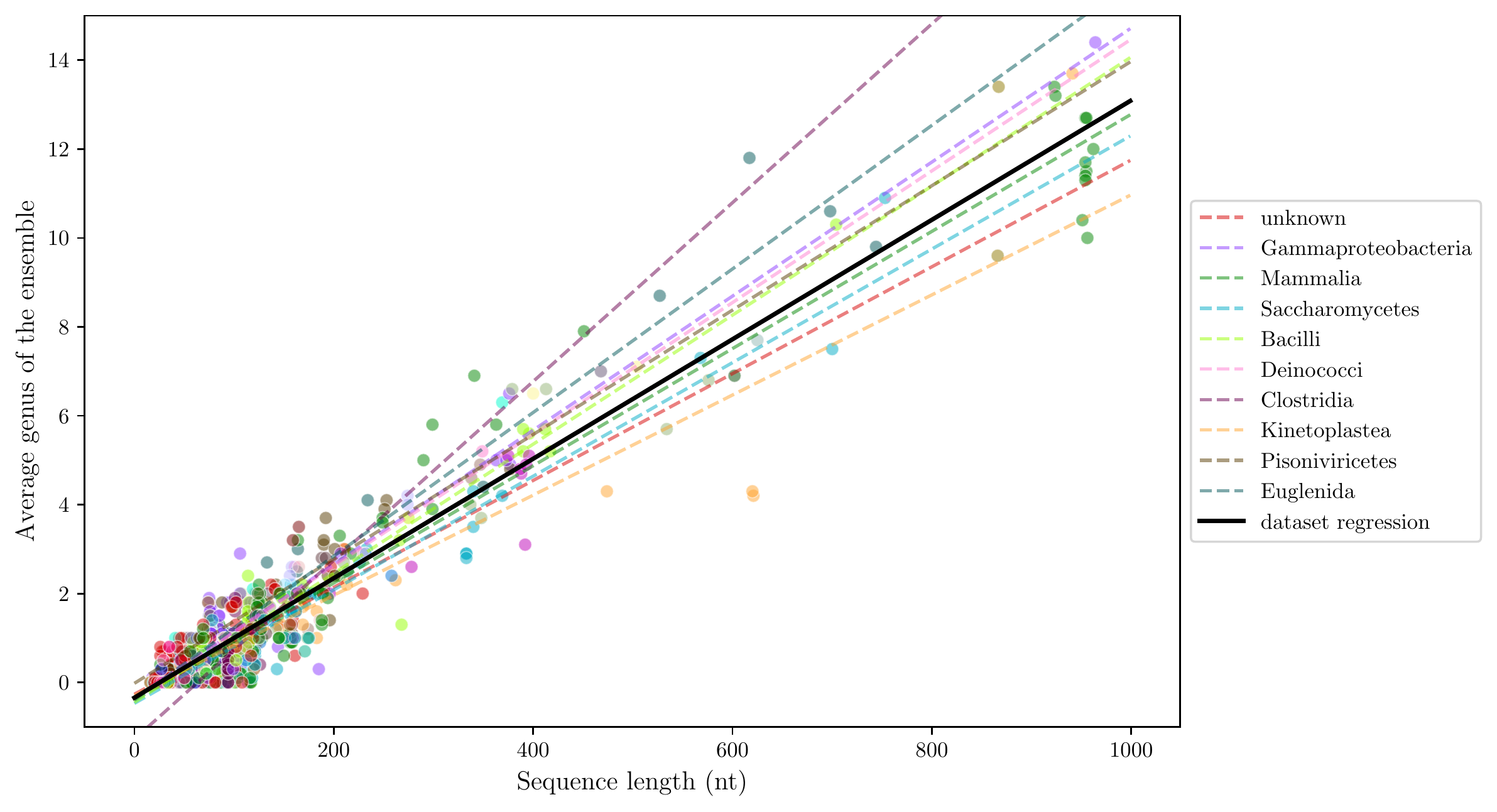}}    
        \caption{Scatter plots and regression lines of how the predicted average genus compares with the sequences length. The two plots show the exact same data and only highlight differences in distinct types of RNA (\textbf{a}) and taxonomic classes (\textbf{b}). In (\textbf{a}) the `dataset regression' line equation and $r^2$ values are shown; the line is also re-drawn in (\textbf{b}) for better comparison. Regression lines for distinct types of RNA and taxonomic classes were only drawn for the most populated groups; for the sake of space, in (\textbf{b}), only the most populated groups are reported in the legend.}
        \label{fig:linreg}
\end{figure*}

Because we were interested in understanding the possible biological implications of the genus, we explored the differences in distinct types of RNA molecules and taxonomic classes (see Fig. \ref{fig:linreg} (\textbf{a}) and (\textbf{b})).

Interestingly, it does not seem that different classes of RNA molecules behave in different ways (see Fig. \ref{fig:linreg} (\textbf{a})). Yet, it has to be remarked that rRNAs, due to their average high length, have been particularly penalized in terms of discarded sequences. Thus, it would be interesting to see if this behaviour would persist in case of their consideration. Surprisingly, it does not seem that enzymatically active RNA molecules, like the rybozymes, possess a higher or lower genus compared to more structural RNA molecules like the tRNA, which have a more `structural' role.

On the other hand, looking at the most represented taxonomic classes, there seems to be a more pronounced difference (see Fig. \ref{fig:linreg} (\textbf{b})), with some classes having a more steep genus/length relationship. Due to the intrinsic biases present in the dataset used, and since not all taxonomic classes are equally and homogeneously represented, it is hard to draw definitive conclusions.

Yet, despite no clear pattern is visible, as the three domains of life display an indistinguishable behaviour, it would be interesting to understand weather other factors, \textit{i.e.}\ environmental conditions, can play a role. Considering there is an intimate relationship between genus and energy, it could be unsurprising to find, for instance, extermophyles to possess an averagely higher genus. It has been already demonstrated, in fact, that RNA structure can be phylognetically conserved even when sequences are not \cite{CaetanoAnolles2002, Hochsmann2004,Wuyts2000}. To test this hypothesis, and assess weather McGenus could detect such differences, 3D structures of homologs of the same RNA molecules should be computationally and experimentally investigated. To this regard, it is important to note that in \cite{Zajac2018}, authors able to assign different total genus values to the same sequence represented in different 3D structures. This is not surprising, since different experimental methodologies can favor different structures and bonds. Unfortunately, McGenus is not able to take into account different conditions, such as, for instance, the presence of ions. These parameters would pose an invaluable addition if integrated into the algorithms presently underlying McGenus. 



\section{Conclusions and future perspectives}
\label{Sec5}

In this study we recover a useful algorithm for predicting RNA structures based on the computation of pseudoknots, and we urge to investigate the biological and technical implications of the concept of genus. After a brief overview of the theoretic background, originally coming from quantum field theory, a comparison  between two approaches to calculate the genus of RNA molecules, \textit{i.e.}\ an experimentally-based one and an \textit{ab-initio} one, was performed. 
Lastly, some basic relationships were investigated, which results are extensively reported in Sect.\ \ref{Sec4}. 

McGenus is still an interesting algorithm with a significant potential, deserving a suitable update in computational and functional terms, now possible thanks to the last decade of advancements in computer science. Moreover, the newest generalizations of the concept of genus, such as genus trace and genus fingerprint matrix, quoted in Sect.\ \ref{Sec3}, could also be integrated as we proved in Sect.\ \ref{Mc_results}. This would allow for the prediction of such functions and matrices, while treating the RNA molecule as a statistical ensemble, potentially providing more information than the genus trace of a crystalline and non-dynamic structure. 

The results outlined in Sect.\ \ref{Sec42} deserve further analyses, as the preliminary tests show an intriguing behaviour within different taxonomic classes. Previous works proved a relationship between RNA structure and biological function \cite{CaetanoAnolles2002, Hochsmann2004,Luwanski2022, Quadrini2019,Wuyts2000}, thus similar questions apply to the concept of genus. This should be seen as the first step toward a new research line, which could lead to a more complete understanding of the relationship between the concept of genus and its biological significance, thus improving our comprehension of RNA 2D and 3D structure and its related functions. 

In the future, in order to exploit engineered RNA molecules, it will be crucial to be able to predict complex 3D structures starting only from the respective sequences. As briefly outlined in Sect.\ \ref{Sec1}, current methods for either tertiary structure prediction or genus calculation need experimental structural data either to train a \ac{ML} algorithm \cite{Dykstra2022,Townshend2021,Zhang2022} or for the computation itself \cite{Luwanski2022,Zajac2018}. Having a physics-based, \textit{i.e.\ ab-initio}, prediction algorithm would be of great assistance in both cases, thus reducing the needs for large experimental campaigns, now needed to collect the necessary data, thus fostering the development of RNA molecules applications. 

Lastly, considering that the algorithm, and cognate theory, underlying McGenus are based on thermo-dynamical properties, it should be possible to expand it to take into account also non-canonical nucleotides and, perhaps, specific environmental conditions. Similarly to the introduction of the concept of genus trace \cite{Rubach2019}, further exploiting the graph theory could yield additional properties of RNA molecules associated with their respective predictions, although this has yet been demonstrated. As a polymer of nucleotides, also RNA molecules are able to store and carry information, and it can be hypothesized that a fraction of such an information can be expressed by genus and its generalizations. It is thus of interest to understand the presence of a the link between the genus and information theory. 

Mathematical tools are proving to be essential to the study of life sciences. Thus to avoid a slowdown of achievements and results, it is crucial for biological researchers and engineers to investigate and evaluate theories and techniques arising in different domains of science \cite{Park2023}. Thus, an update of the algorithm, and the software, based on well-studied mathematical formalism, is compelling, as well as the continuous research and development of computational methods borrowed from different scientific realms, as for instance shown in the present paper, from particle physics theories. 
\end{multicols}

\section*{Data availability} 
The full dataset is available as tab separated file at: \url{https://github.com/grassoste/genus-comparisons}.





\paragraph{Acknowledgements.}
The authors are very grateful to Graziano Vernizzi for the initial support and the exhaustive introduction to the world of genus diagrams. Moreover, they would like to thank Michaël Bon for providing us the original McGenus code.

NC is member and acknowledge the support of
Gruppo Nazionale per l’Analisi Matematica, la Probabilità e le loro Applicazioni
(GNAMPA) of Istituto Nazionale di Alta Matematica (INdAM). Moreover, NC is supported by the MIUR - PRIN 2017 project “From
Models to Decisions” (Prot. N. 201743F9YE).

\bibliographystyle{abbrv} 
\bibliography{ref.bib}

\begin{thebibliography}{10}

\bibitem{Andersen13a}
J.~E. Andersen, L.~O. Chekhov, R.~C. Penner, C.~M. Reidys, and P.~Sułkowski.
\newblock Topological recursion for chord diagrams, {RNA} complexes, and cells
  in moduli spaces.
\newblock {\em Nucl. Phys. B}, 866(3):414--443, 2013.

\bibitem{Andersen13b}
J.~E. Andersen, R.~C. Penner, C.~M. Reidys, and M.~S. Waterman.
\newblock Topological classification and enumeration of {RNA} structures by
  genus.
\newblock {\em J. Math. Biol.}, 67(5):1261--1278, 2013.

\bibitem{Berman2000a}
H.~M. Berman, J.~Westbrook, Z.~Feng, G.~Gilliland, T.~N. Bhat, H.~Weissig,
  I.~N. Shindyalov, and P.~E. Bourne.
\newblock {The Protein Data Bank}.
\newblock {\em Nucleic Acids Res.}, 28(1):235--242, 2000.

\bibitem{Bon12}
M.~Bon, C.~Micheletti, and H.~Orland.
\newblock {{M}c{G}enus: a {M}onte {C}arlo algorithm to predict {RNA} secondary
  structures with pseudoknots}.
\newblock {\em Nucleic Acids Res.}, 41(3):1895--1900, 2012.

\bibitem{Bon10}
M.~Bon and H.~Orland.
\newblock Prediction of {RNA} secondary structures with pseudoknots.
\newblock {\em Phys. A: Stat. Mech. Appl.}, 389(15):2987--2992, 2010.

\bibitem{Bon11}
M.~Bon and H.~Orland.
\newblock {{TT2NE}: a novel algorithm to predict {RNA} secondary structures
  with pseudoknots}.
\newblock {\em Nucleic Acids Res.}, 39(14):e93, 2011.

\bibitem{Bon08}
M.~Bon, G.~Vernizzi, H.~Orland, and A.~Zee.
\newblock Topological classification of {RNA} structures.
\newblock {\em J. Mol. Biol.}, 379(4):900--911, 2008.

\bibitem{Bouttier11}
J.~Bouttier.
\newblock Matrix integrals and enumeration of maps.
\newblock In {\em The Oxford Handbook of Random Matrix Theory}, chapter~26.
  Oxford University Press, Oxford, UK, 2011.

\bibitem{Burns2015}
J.~Burns, N.~Jonoska, and M.~Saito.
\newblock Genus ranges of chord diagrams.
\newblock {\em J. Knot Theory Ramif.}, 24(4), 2015.

\bibitem{CaetanoAnolles2002}
G.~Caetano-Anolles.
\newblock Tracing the evolution of {RNA} structure in ribosomes.
\newblock {\em Nucleic Acids Res.}, 30(11):2575--2587, 2002.

\bibitem{Sweeney2020}
R.~Consortium.
\newblock {RNAcentral 2021: secondary structure integration, improved sequence
  search and new member databases}.
\newblock {\em Nucleic Acids Res.}, 49(D1):D212--D220, 2020.

\bibitem{Dam1992}
E.~T. Dam, K.~Pleij, and D.~Draper.
\newblock Structural and functional aspects of rna pseudoknots.
\newblock {\em Biochemistry}, 31(47):11665--11676, 1992.

\bibitem{Denay2021}
G.~Denay.
\newblock Cvua-rrw/taxidtools: 2.2.3, {DOI}:
  \href{https://zenodo.org/record/5556006#.ZC0-eXZBw2w}{10.5281/zenodo.5556006},
  2021.

\bibitem{Dykstra2022}
P.~B. Dykstra, M.~Kaplan, and C.~D. Smolke.
\newblock Engineering synthetic {RNA} devices for cell control.
\newblock {\em Nat. Rev. Genet.}, 23(4):215--228, 2022.

\bibitem{Fallmann2017}
J.~Fallmann, S.~Will, J.~Engelhardt, B.~Grüning, R.~Backofen, and P.~F.
  Stadler.
\newblock Recent advances in {RNA} folding.
\newblock {\em J. Biotechnol.}, 261:97--104, 2017.

\bibitem{Fudenberg2012}
G.~Fudenberg and L.~A. Mirny.
\newblock Higher-order chromatin structure: bridging physics and biology.
\newblock {\em Curr. Opin. Genet. Dev.}, 22(2):115--124, 2012.

\bibitem{Hochsmann2004}
M.~Hochsmann, B.~Voss, and R.~Giegerich.
\newblock Pure multiple {RNA} secondary structure alignments: a progressive
  profile approach.
\newblock {\em {IEEE}/{ACM} Trans. Comput. Biol. Bioinform.}, 1(1):53--62,
  2004.

\bibitem{Hooft74}
G.~Hooft.
\newblock A planar diagram theory for strong interactions.
\newblock {\em Nucl. Phys. B}, 72(3):461--473, 1974.

\bibitem{Karniadakis2021}
G.~E. Karniadakis, I.~G. Kevrekidis, L.~Lu, P.~Perdikaris, S.~Wang, and
  L.~Yang.
\newblock Physics-informed machine learning.
\newblock {\em Nat. Rev. Phys.}, 3(6):422--440, 2021.

\bibitem{Luwanski2022}
K.~Luwanski, V.~Hlushchenko, M.~Popenda, T.~Zok, J.~Sarzynska, D.~Martsich,
  M.~Szachniuk, and M.~Antczak.
\newblock {RNAspider}: a webserver to analyze entanglements in {RNA} 3{D}
  structures.
\newblock {\em Nucleic Acids Res.}, 50(W1):W663--W669, 2022.

\bibitem{Madeira2022}
F.~Madeira, M.~Pearce, A.~R.~N. Tivey, P.~Basutkar, J.~Lee, O.~Edbali,
  N.~Madhusoodanan, A.~Kolesnikov, and R.~Lopez.
\newblock Search and sequence analysis tools services from {EMBL}-{EBI} in
  2022.
\newblock {\em Nucleic Acids Res.}, 50(W1):W276--W279, 2022.

\bibitem{Orland2002}
H.~Orland and A.~Zee.
\newblock {RNA} folding and large {$N$} matrix theory.
\newblock {\em Nucl. Phys. B}, 620(3):456--476, 2002.

\bibitem{Park2023}
M.~Park, E.~Leahey, and R.~J. Funk.
\newblock Papers and patents are becoming less disruptive over time.
\newblock {\em Nature}, 613(7942):138--144, 2023.

\bibitem{Pillsbury05}
M.~Pillsbury, H.~Orland, and A.~Zee.
\newblock Steepest descent calculation of {RNA} pseudoknots.
\newblock {\em Phys. Rev. E}, 72:011911, 2005.

\bibitem{Quadrini2019}
M.~Quadrini, L.~Tesei, and E.~Merelli.
\newblock An algebraic language for {RNA} pseudoknots comparison.
\newblock {\em {BMC} Bioinform.}, 20(S4), 2019.

\bibitem{Rubach2019}
P.~Rubach, S.~Zajac, B.~Jastrzebski, J.~I. Sulkowska, and P.~Su{\l}kowski.
\newblock Genus for biomolecules.
\newblock {\em Nucleic Acids Res.}, 48(D1):D1129--D1135, 2019.

\bibitem{Serra95}
M.~J. Serra and D.~H. Turner.
\newblock Predicting thermodynamic properties of {RNA}.
\newblock {\em Meth. Enzymol.}, 259:242--261, 1995.

\bibitem{Townshend2021}
R.~J.~L. Townshend, S.~Eismann, A.~M. Watkins, R.~Rangan, M.~Karelina, R.~Das,
  and R.~O. Dror.
\newblock Geometric deep learning of {RNA} structure.
\newblock {\em Science}, 373(6558):1047--1051, 2021.

\bibitem{vernizzi04pp}
G.~Vernizzi, H.~Orland, and A.~Zee.
\newblock Prediction of {RNA} pseudoknots by {M}onte {C}arlo simulations,
  preprint available on ar{X}iv:
  \href{https://arxiv.org/abs/q-bio/0405014}{q-bio/0405014}, 2004.

\bibitem{Vernizzi04}
G.~Vernizzi, H.~Orland, and A.~Zee.
\newblock Enumeration of {RNA} structures by matrix models.
\newblock {\em Phys. Rev. Lett.}, 94:168103, 2005.

\bibitem{Vernizzi16}
G.~Vernizzi, H.~Orland, and A.~Zee.
\newblock Classification and predictions of {RNA} pseudoknots based on
  topological invariants.
\newblock {\em Phys. Rev. E}, 94:042410, 2016.

\bibitem{Vernizzi06}
G.~Vernizzi, P.~Ribeca, H.~Orland, and A.~Zee.
\newblock Topology of pseudoknotted homopolymers.
\newblock {\em Phys. Rev. E}, 73:031902, 2006.

\bibitem{Weeks2021}
K.~M. Weeks.
\newblock Piercing the fog of the {RNA} structure-ome.
\newblock {\em Science}, 373(6558):964--965, 2021.

\bibitem{Wuyts2000}
J.~Wuyts.
\newblock Comparative analysis of more than 3000 sequences reveals the
  existence of two pseudoknots in area {V}4 of eukaryotic small subunit
  ribosomal {RNA}.
\newblock {\em Nucleic Acids Res.}, 28(23):4698--4708, 2000.

\bibitem{Xu20}
X.~Xu and S.-J. Chen.
\newblock {Topological constraints of {RNA} pseudoknotted and loop-kissing
  motifs: applications to three-dimensional structure prediction}.
\newblock {\em Nucleic Acids Res.}, 48(12):6503--6512, 2020.

\bibitem{Zajac2018}
S.~Zaj{\k{a}}c, C.~Geary, E.~S. Andersen, P.~Dabrowski-Tumanski, J.~I.
  Sulkowska, and P.~Su{\l}kowski.
\newblock Genus trace reveals the topological complexity and domain structure
  of biomolecules.
\newblock {\em Sci. Rep.}, 8(1), 2018.

\bibitem{Zhang2022}
J.~Zhang, Y.~Fei, L.~Sun, and Q.~C. Zhang.
\newblock Advances and opportunities in {RNA} structure experimental
  determination and computational modeling.
\newblock {\em Nature Methods}, 19(10):1193--1207, 2022.

\bibitem{Zuker99}
M.~Zuker, D.~H. Mathews, and D.~H. Turner.
\newblock {\em RNA Biochemistry and Biotechnology}, volume~70 of {\em NATO
  Science Series}, chapter Algorithms and Thermodynamics for RNA Secondary
  Structure Prediction: A Practical Guide.
\newblock J. Barciszewski and B. F. C. Clark (Eds.), Springer, Dordrecht, NL,
  1999.

\end{thebibliography}
\end{document}